**Dissociation slowdown by collective optical response under strong coupling conditions**

Maxim Sukharev,[1,2] Joseph Subotnik,[3] Abraham Nitzan[3]

[1]College of Integrative Sciences and Arts, Arizona State University, Mesa, Arizona 85212, USA
[2]Department of Physics, Arizona State University, Tempe, Arizona 85287, USA
[3]Department of Chemistry, University of Pennsylvania, Philadelphia, Pennsylvania 19104, USA

We consider an ensemble of diatomic molecules resonantly coupled to an optical cavity under strong coupling conditions at normal incidence. Photodissociation dynamics is examined via direct numerical integration of the coupled Maxwell-Schrödinger equations with molecular ro-vibrational degrees of freedom explicitly taken into account. It is shown that the dissociation is significantly affected (slowed down) when the system is driven at its polaritonic frequencies. The observed effect is demonstrated to be of transient nature and has no classical analog. An intuitive explanation of the dissociation slowdown at polaritonic frequencies is proposed.

## 1. Introduction

Spectroscopical and chemical manifestations of strong light-matter coupling are subjects of growing interest[1, 2-4] due to both the underlying fundamental aspects and the promise of interesting applications. In particular, the possibility that the outcome of a photochemical process can be controlled by the geometry and composition of nearby dielectric nanostructures has attracted intense interest. (In this paper we do not consider the possibility that the proximity of metallic interfaces may affect also the outcome of thermal chemical processes without any optical pumping.) Such effects have been suggested[5] and observed[6] in plasmonic cavities, where modification of photochemical process reflects surface-induced local modification of the electromagnetic field and/or the opening of new relaxation channels, as well as the plasmon-induced formation of hot electrons.[7] In more recent studies performed in Fabri-Perot cavity configurations such modifications are small, and photochemistry appears to reflect mostly the excitation of polaritons – hybrid molecule-light states, where strong light-matter coupling is associated with the collective optical response of a large number of molecules.[8, 9, 10]

Polaritonic response of such molecules-optical cavity systems is usually manifested in the observation of Rabi splitting in absorption, emission, scattering, or reflection-transmission of the light, indicating the formation of polaritons – hybrid light-matter states.[11] The collective nature of this interaction is expressed by the linear dependence of this splitting on the square root of the molecular density.[10] The question as to whether this collective response is manifested in other molecular processes in cavity environemnts has been discussed both with respect to vibronic strong coupling phenomena, where cavity mode(s) are strongly coupled to electronic transitions[3, 8, 10, 12-14] and purely vibrational strong coupling where the cavity resonates with vibrational motions.[3, 9, 10, 15] Here, we focus on the former class of phenomena, noting that while theories associated with the behavior of a few (1-3) molecules are clear,[10, 16, 17] strong coupling in Fabri-Perot configurations arises from collective molecular response. Theoretical treatments of such situations[14, 16, 18-21] have heretofore applied the the Holstein-Tavis-Cummings (HTC) model

$$\hat{H}_{HTC} = \hbar\omega_c \hat{a}^\dagger \hat{a} + \hbar \sum_{j=1}^{N} \left[ \omega_{xg} \hat{\sigma}_j^+ \hat{\sigma}_j^- + \frac{g}{2}(\hat{a}^\dagger \hat{\sigma}_j^- + \hat{\sigma}_j^+ \hat{a}) + \omega_v \hat{b}_j^\dagger \hat{b}_j + \lambda \hat{\sigma}_j^+ \hat{\sigma}_j^- (\hat{b}_j^\dagger + \hat{b}_j) \right] \quad (1)$$

which combines the Tavis-Cummings (TC) model of $N$ 2-level atoms interacting with an electromagnetic field mode[22] with the Holstein polaron model[23] to include coupling of a  molecular electronic transition with a cavity mode on the one hand and a molecular (harmonic) vibration on the other. In Eq. (1), each molecules is a two-electronic-state entity and (in the minimal version of the model) a single exciton case is considered and direct intermolecular interactions are ignored. The operator $\hat{a}$ ( $\hat{a}^{\dagger}$) annihilates (creates) a photon of a cavity mode of frequency $\omega_c$ while $\hat{\sigma}_j = |g_j\rangle\langle e_j|$ and $\hat{\sigma}_j^{\dagger} = |e_j\rangle\langle g_j|$ respectively affect the upward and downward transitions between the lower $|g_j\rangle$ and upper $|e_j\rangle$ electronic states of molecule $j$, $\hbar\omega_{xg}$ is the molecular electronic transition energy and $g$ is the molecule-cavity mode interaction matrix element, taken to be the same for all molecules. The above are parameters of the TC Hamiltonian. In addition, each molecule is associated with one harmonic oscillator of frequency $\omega_v$, described (for molecule $j$) by the raising and lowering operators $\hat{b}_j^{\dagger}$ ($\hat{b}_j$) and vibronic interaction that corresponds to a shift of the molecular nuclear harmonic surfaces between the ground and excited electronic states and quantified by the magnitude λ of this shift.

The HTC model, Eq. (1), accounts well for the main spectroscopic observations: The formation of hibride molecule/cavity-photon states that appear as a Rabi-split signal in the linear optical response of such systems as well as the characteristc dependence of the splitting, $\Omega_R = g\sqrt{N}$, on the numebr of molecules $N$ (g characterizing the coupling between individial molecules and the cavity photon). The implication for the nuclear dynamics following such excitations has been discussed by several groups,[4, 12-14, 16, 20, 21, 24] leading to interesting predictions about the possible implication of strong coupling to a cavity mode on the standard Born-Oppenheimer (BO) picture of molecular nuclear dynamics. Arguments involving timescale speration between fast (electronic and cavity modes) and slow (vibrational) motions have led Spano,[20] and later Herrera and Spano[14] to suggest that in the large molecular number $(N)$ limit, the nuclear potential surface associated with a polariton state is similar (up to deviations of order $N^{-1}$) to the ground state potential surface, essentially implying vibronic decoupling with strong implications for subsequent interstate electronic transitions including electron transfer processes. Later theoretical studies[4, 16, 25, 26] have indeed confirmed the existence of vibronic decoupling, however in a restrictive form: provided that the Rabi splitting is large relative to the reorganization energy associated with the vibronic coupling. It has been suggested that such vibronic decoupling may have dramatic consequences for photophysical and photochemical processes initiated by polariton excitation.[12, 14, 18, 26, 27, 28] Another effect stems from the polariton shift, which can place the excitation energy closer or further away from reactive channels. Experimenhtally, predictions based on vibronic (or polaron) decoupling have not yet been verified experimentally, while effects associated with polaritonic energy shift have been observed by several workers.[28, 29]

Further efforts to bridge between theory and experiments need to consider some shortcomings of the HTC mode as usually applied: First, while the model of Eq. (1) considers a single (undamped) cavity mode, assuming near resonance between the molecular electronic transition and this modes, realistic cavity involves many modes whose dynamics should reflect also the way by which the external field interacts with the cavity. A full description of experimental observables should have the capacity to describe the way by which the incident field generates transmission and reflection signals as well as energy dissipation in the mirror environments.[1] Second, theoretical considerations based on Eq. (1) have been limited to considering the dynamics following polariton excitation in the single exciton subspace, where one photon absorbed from (and later emitted to) the far field forms a superposition of states in which only one molecule is excited or the cavity mode occupies its lowest excited level while higher order excitations of our system are disregarded. This approximation appears to be valid for weak incident fields, however its actual implications are presently unknown. Third, nuclear motion in the model (1) is represented by a harmonic interaction and vibronic coupling is associaed with a shift in the equilibrium position that defines the harmonic nuclear potential. Consquently, the ensuing dynamics cannot describe a reactive processes such as molecular photodissociation.

Finally, while arguments concerning timescale separations are instrumental for understading molecular vibronic dynamics, and for a cavity photon in resonance with amolecular electronic transition, a picture of fast electronic-photonic dynamics vs slow nuclear dynamics is by itself valid, it is important to note that in Eq. (1) the electronic dynamics is dominated by two different timescales. The timescale associated with the Rabi splitting, $\Omega_R = g\sqrt{N}$ can be of order $\tau_R = \Omega_R^{-1}$ : $10$ fs or even shorter [18], much smaller than characteristic molecular nuclear time $\tau_{Nuc}$. That being said, the timescale associated with energy transfer between molecules is of order $\tau_{ET} = \tau_R \sqrt{N} = g^{-1}$ where $N$ can be of order $10^5$ as a conservative estimate.[30] These timescales thus satisfy

$$\tau_R < \tau_{Nuc} < \tau_{ET}, \tag{2}$$

Thus, the implications of the timescale separation between the cavity-coupled molecular electronic dynamics and the molecular nuclear dynamics should be scrutinized more carefully.

In this paper we introduce a semiclassical simulation model that address the above issues at the cost of other significant approximations whose effect need to be carefully addressed as well. In this model, the molecules are represented, as in the TC model, by two electronic state and (in the present implementation) one nuclear mode. The assumption that the nuclear motion is harmonic is relaxed, so reactive potential suraces may be considered and in the implementation described below the upper potential surface is taken to be repulsive. The cavity is described as a metalic structure of any desired shape where the metal is described by its dielectric response function, and is taken here to comprise two parallel metal plates of a given thickness with the molecules distributed in the space between them. Solving the corresponding Maxwell-Schrödinger equation for this system yields the full scattering dynamics and in particular the transmission and reflection signals obtained in response to an incident light. The frequency dependence of these signals reflects the consquences of light-molecule coupling in the presence of the metal interfaces and, as shown below, account well for the polaritonic nature of the calculated optical respnse. At the same time, the reactive dynamics following molecular exitation can be numerically monitored and its behavior in and out of the cavity can be characterized.

Details of our model and our computational procedure, as well as the approximation involved are provided and discussed in Sect. 2. Results of our calculations are presented, compared and discussed in Section 3. Section 4 concludes.

---

[1] Admittedly, a full description of the radiation field should also take into account the intermolecular interactions that are diregarded in (1), however the model considrered in this paper still falls short of this goal as it considers only the transverse part of the EM field.

## 2. Model and computational procedure

We consider a general problem of quantum emitters driven by a classical electromagnetic (EM) radiation comprising the incident field and the scattered field at a metallo-dielectric interface of arbitrary shape. Electrodynamics follows Maxwell's equations

$$\frac{\partial \vec{B}}{\partial t} = -\nabla \times \vec{E},$$
$$\frac{\partial \vec{E}}{\partial t} = c^2 \, \nabla \times \vec{B} - \frac{1}{\varepsilon_0} \frac{\partial \vec{P}}{\partial t}, \tag{3}$$

where the macroscopic polarization, $\vec{P}$, accounts for material response.

In this work, the molecular quantum emitters are modeled using a two-state diatomic model with ro-vibrational degrees of freedom. For each molecule we employ the Born-Oppenheimer expansion of the electro-nuclear wave function, $\Psi\left(\vec{r}, \vec{R}, t\right)$, separating electronic, $\vec{r}$, from nuclear, $\vec{R}$, motion

$$\Psi\left(\vec{r}, \vec{R}, t\right) = \chi_g\left(\vec{R}, t\right) \varphi_g\left(\vec{r}, \vec{R}\right) + \chi_e\left(\vec{R}, t\right) \varphi_e\left(\vec{r}, \vec{R}\right), \tag{4}$$

where $\varphi_{g,e}$ are the electronic wavefunctions of the ground/excited electronic state, respectively, and $\chi_{g,e}$ are the time-dependent ro-vibrational wavefunctions.

The rotational degree of freedom is taken into account by expanding the nuclear wavefunctions in the basis set of normalized Wigner rotational matrices $D_{M,0}^J$ with $J$ and $M$ denoting the molecular rotational quantum number and its projection on a fixed axis of the laboratory frame. The expansion reads[31]

$$\chi_{g,e}\left(\vec{R}, t\right) = \sum_{J,M} \xi_{JM}^{(g,e)}\left(R, t\right) D_{M,0}^J\left(\hat{R}\right). \tag{5}$$

Here we use the following notation for the nuclear coordinate $\vec{R} \equiv \left(R, \hat{R}\right)$ to emphasize the angular dependance of the Wigner functions. The expansion (5) is plugged into the time-dependent Schrödinger equation with a subsequent projection onto the electronic and rotational basis functions. This yields a set of coupled differential equations defining the ro-vibrational dynamics of $\xi_{JM}^{(g,e)}\left(R, t\right)$ on two potential energy surfaces

$$i\hbar \frac{\partial \xi_{JM}^{(g)}}{\partial t} = \hat{H}_J^{(g)} \xi_{JM}^{(g)} - d_{ge}\left(R\right) \sum_{J'M'} \left(E_x X_{JM}^{J'M'} + E_y Y_{JM}^{J'M'} + E_z Z_{JM}^{J'M'}\right) \xi_{J'M'}^{(e)},$$
$$i\hbar \frac{\partial \xi_{JM}^{(e)}}{\partial t} = \hat{H}_J^{(e)} \xi_{JM}^{(e)} - d_{eg}\left(R\right) \sum_{J'M'} \left(E_x X_{JM}^{J'M'} + E_y Y_{JM}^{J'M'} + E_z Z_{JM}^{J'M'}\right)^* \xi_{J'M'}^{(g)}, \tag{6}$$

where $d_{ge}\left(R\right)$ is the transition dipole. The matrix elements $X/Y/Z_{JM}^{J'M'}$ correspond to projecting the local electric field, $\vec{E}$, that follows (3) in laboratory frame to the molecular frame. The general expressions for the matrix elements can be found in Ref [32] and are expressed using 3$j$-symbols. The effective Hamiltonians in (6) read

$$\hat{H}_J^{(g,e)} = -\frac{\hbar^2}{2\mu}\left(\frac{\partial^2}{\partial R^2} - \frac{J\left(J+1\right)}{R^2}\right) + V_{g,e}\left(R\right), \tag{7}$$

where $\mu$ is the reduced molecular mass and $V_{g,e}$ are the electronic potential energy surfaces. The induced dipole moment, $\vec{d}$, of each molecule is evaluated and the macroscopic polarization is computed following the mean field approximation in accordance to

$$\vec{P} = N_M \, \vec{d}, \qquad (8)$$

where $N_M$ is the number density of molecules. It is coupled back to Maxwell's equations (3) via polarization current. The complete model employed in this work relies on the system of coupled partial differential equations (3) and (6). We note that equations (6) are driven by a local electric field, which in principle depends sensitively on spatial coordinates. We thus need to propagate quantum equations at every grid point and couple their dynamics to Maxwell's equations.

Numerically such a propagation is achieved via combining finite-difference time-domain (FDTD) methodology for Maxwell's equations (3) with the split-operator propagation of the Schrodinger equations (6). In general, full three-dimensional calculations are extremely memory and time consuming. While three dimensional calculations can be made more efficient by 3D domain decomposition parallelization,[33] this work considers a one-dimensional optical cavity as schematically depicted in Fig. 1a. The cavity is formed by two parallel golden plates. The electrodynamics of the cavity follows the Maxwell equations (3) that, for the chosen geometry, reduce to

$$\frac{\partial B_y}{\partial t} = -\frac{\partial E_x}{\partial z},$$
$$\frac{\partial E_x}{\partial t} = -c^2 \frac{\partial B_y}{\partial z} - \frac{1}{\varepsilon_0} \frac{\partial P_x}{\partial t}, \qquad (9)$$

where it is assumed that the cavity size is significantly smaller compared to its transverse directions (i.e., all spatial dependencies of the electromagnetic field with respect to $x$ and $y$ are neglected). The macroscopic polarization $P_x$ inside the metal is simulated using the conventional Drude model[34]

$$\frac{\partial^2 P_x}{\partial t^2} + \gamma \frac{\partial P_x}{\partial t} = \varepsilon_0 \, \Omega_p^2 \, E_x, \qquad (10)$$

with parameters describing gold: $\gamma = 0.181$ eV and $\Omega_p = 7.04$ eV. The molecular part of the model (6) retains only $E_x$ and the corresponding matrix elements $X_{JM}^{J'M'}$.[35]

Results reported in this manuscript pertain to the cavity driven by the CW field at normal incidence propagating along $z$ from right to left, i.e., exciting the right metal plate first. By adjusting plates' thickness and the cavity size (distance between plates) we can tune the fundamental cavity mode to a desired frequency. We note that the thickness of the plates controls how much electromagnetic energy can penetrate the cavity and remain inside at resonant conditions. The thickness also influences the cavity losses, which include two channels – radiation to the far field and Ohmic losses in the metal. Eqs. (9) and (10) are discretized in space and time following the FDTD approach. To simulate an open system, we enclose the simulation domain with artificial absorbing boundaries via the convolutional perfectly matched layers (CPML) technique with 20 layers on each side.[36] The latter corresponds to the artificial reflection coefficient of $10^{-12}$ ensuring the numerical stability. In the present implementation, to achieve numerical convergence the spatial resolution was set at $\delta z = 1$ nm with a time step $\delta t = \delta z / 2c$. The entire FDTD grid was taken 2 μm wide with the cavity placed in its center.

The molecular slab is placed at the center of the cavity as shown in Fig. 1a. The transition dipole, $d_{eg} = d_{ge}$, is assumed to be independent from $R$ and is set to 10 Debye in all calculations. The ground electronic state is taken in the form of a Morse function

$$V_g(R) = D_g \left(1 - \exp\left(-a_g\left(R - R_g\right)\right)\right)^2 - D_g,$$ (11)

while the excited, dissociative electronic state reads

$$V_e(R) = 2D_e \exp\left(-a_e\left(R - R_e\right)\right).$$ (12)

In the calculations presented below we have used the following set of parameters (all in atomic units) corresponding to Li$_2$: $D_g = 3.86 \times 10^{-2}$, $a_g = 0.458$, and $R_g = 5.06$, $D_e = 1.77 \times 10^{-2}$, $a_g = 0.364$, and $R_g = 3.98$.[31]

The resulting equations were numerically propagated using the split-operator method[37] on the grid $R = $ (3 a.u. – 100 a.u.) with the total of 1024 points. In all simulations we explicitly check whether $\xi_{JM}^{(e)}$ reaches the end of the grid. Such an occurrence would correspond to a spurious appearance of the part of the wavepacket at 3 a.u. (since the split-operator method employs Fast Fourier Transform with periodic boundary conditions) and in turn unphysically distort the nuclear dynamics. We performed several numerical experiments and found that for our choice of parameters, if the $R$-grid is terminated at 100 a.u., simulations can safely be performed up to 500 fs. The dipole moment, $d_x$, of each molecule is evaluated at every time step and the macroscopic polarization is computed following the mean field approximation[38]

$$P_x = N_M d_x,$$ (13)

where $N_M$ is the number density of molecules in the slab. The polarization is coupled back to Maxwell's equations (9) via polarization current (last term in Eq. (9b)). The resulting system comprised of Eqs. (9), (10), and (6) is solved numerically as described above.

Molecules are initially in the ground electronic state occupying the ground ro-vibrational state. The process starts when molecules are subjected to a low intensity CW that drives the ground ro-vibrational state into the dissociative potential as schematically depicted in Fig. 1b. The range of transition frequencies corresponding to such a transition are defined by the Frank-Condon (FC) factor

$$FC = \left|\left\langle \xi_{1,0}^{(e)}(E) \middle| \xi_{0,0}^{(g)}(v=0) \right\rangle\right|^2.$$ (14)

The FC distribution for the numerical parameters used is shown in Fig. 1c. This makes it simple to find a desired resonant cavity (here chosen to match the maximum of the FC distribution) by adjusting mirror-to-mirror distance. The results of simulations combining the cavity and molecules are shown in Fig. 1d that shows the transmission through an empty cavity and a cavity containing the molecular slab for several values of the number density, $N_M$. The observed Rabi splitting indicates that the system is in the strong coupling regime. We also note that this splitting follows expected $\sqrt{N_M}$ dependence.[2] Although we consider the cavity mode that is in nearly perfect resonance with a frequency corresponding to the maximum of the Frank-Condon distribution (Fig. 1c), clearly the upper polariton has a higher population as it becomes more evident at high molecular concentrations. This can be explained by the material dispersion of the cavity mode (the linear transmission shown as black line in Fig. 1d), which is not symmetric for the frequencies above and below the fundamental frequency.

It is informative to examine the local field enhancement (relative to the incident field amplitude) inside the cavity as a function of field carrier frequency and the longitudinal coordinate $z$. Fig. 2a compares the ensemble average of the local field enhancement with the linear transmission presenting them as functions of the driving frequency. The local field is enhanced at the polaritonic frequencies and the lower and upper polariton frequencies seen in the enhancement spectrum are nearly identical to those seen in the transmission signal. Additionally, the ratio of the upper to the lower polariton contributions to the local field and transmission are noticeably different: 1.53 for the field enhancement and 1.92 for the transmission. This indicates that the cavity mode dispersion plays less of a critical role for the local field that drives molecules. Furthermore, the spatial dependence of the field enhancement vs carrier frequency is shown in Figs. 2b and 2c. The cavity mode is localized in the center of the cavity peaking at 1.69 eV. When the molecular slab is present and the cavity is tuned to the transition frequency, two polaritonic states are observed. Both exhibit spatial localization similar to the fundamental mode in Fig. 2b.

### 3. Results: Molecular photodissociation under strong coupling conditions

We now examine how the system reaches the steady-state regime after the onset of an incident field amplitude with a rise time of 5 fs, during which the pump vertically projects the ground vibrational state onto the excited electronic potential energy surface. After propagating the coupled Maxwell-Schrödinger equations (9) and (6), we find that the system quickly reaches a plateau under these effectively CW conditions In Fig. 3a, we study the resulting ensemble average nuclear dynamics in the excited electronic state and plot $\left| \chi_e \left( \mathbf{R}, t \right) \right|^2$ over the first 250 fs. At short times, the peak in the nuclear wavepacket appears at 5 a.u. indicating the Frank-Condon transition. At later times, the peak remains at the same coordinate due to continuous pumping from the ground electronic state, but the tail begins to increase to larger distances as the wavefunction spreads over the dissociative potential surface. As a practical definition, we regard the system as reaching "steady state" when the R probability flux at 30 a.u. defined as

$$j_R = \frac{\hbar}{\mu} \mathrm{Im} \left( \chi_e^* \frac{\partial \chi_e}{\partial R} \right). \tag{15}$$

no longer depends on time. For the range of parameters considered, this situation is reached at about 450 fs.

Next, Fig. 3b shows the probability flux in the excited electronic state as a function of time at three driving frequencies: lower polariton (black), central transition frequency (red), and the upper polariton (blue). Calculations are performed at $R = 15$ a.u which is reached by the excited nuclear wavepacket after over 100 fs. We clearly see that the probability flux exhibits noticeably different rise times, which depend on the driving frequency. When this frequency is resonant with the bare molecular transition frequency (red line in Fig. 3b), the wavefunction develops a large wavefront at early times, which moves to larger distances (as seen in Fig. 3a which is also calculated for the molecular transition frequency). The flux peaks at about 180 fs and then becomes smaller. When driving at the upper/lower polaritonic frequencies the propagation is different from the central frequency excitation. To better quantify this behavior, we calculate the running time integral of the probability flux and use the linear regression to extract the corresponding slope and intercept. Their ratio corresponds to the average rise time. Results of these calculations are shown in Fig.

3c for molecules in and out of the resonant cavity. By this measure, it is seen the dissociative dynamics is significantly slowed down when the driving frequency is resonant with either lower or upper polaritons.

To examine this phenomenon from a different perspective, we calculate the average distance traveled by the nuclear motion at a relatively long time, after "steady state" (as defined above) is achieved. Specifically, we calculate the average internuclear separation of excited molecules, i.e., the weighted average internuclear separation given by

$$R_e = \frac{\langle \chi_e | R | \chi_e \rangle}{\langle \chi_e | \chi_e \rangle}. \tag{16}$$

Following the initial switch-on, the system is driven by the CW field until we reach the "steady-state" where the calculated probability flux at large R (here taken at $R = 50$ a.u.) is constant corresponding to 450 fs. At this time the internuclear distance in the excited state, Eq. (16), is calculated for different values of the driving frequency and for different molecular densities. The results are shown in Fig. 4 together with the corresponding linear transmission spectra. When the molecular slab is in vacuum, $R_e$ varies linearly with the pump frequency (black line in Fig. 4) and is nearly independent of the molecular density. When the molecules are placed inside a resonant cavity, the dynamics drastically change demonstrating that the nuclear wavepacket is significantly slower near the polaritonic frequencies.

Some indication regarding the origin of the observed phenomenon is shown in Fig. 5. Here we examine the CW dynamics of $R_e$ (panel (a)) at different driving frequencies. Once the CW drive is established, the system reaches the steady state regime and $R_e$ becomes a linear function of time meaning that the dissociation rate is constant. It is readily seen from Fig. 5a that such a rate is frequency dependent. Next, we perform a linear regression of $R_e$ once the system reaches the steady state, and we evaluate the slope and the intercept as functions of the driving frequency (panel (b)). The slope (dashed black line) does not depend on the number density of molecules and varies approximately linearly with the frequency. The intercept, however, exhibits significant changes showing two well-resolved minima at polariton frequencies. In agreement with the results of Fig. 4b this again illustrates that the slowdown of the dissociation process when the optical pumping is done at the polariton frequency is an *early* time phenomenon.

## 4. Discussion

The results described in Section 3 were obtained using two important approximations. First, the electrodynamics field is treated classically, and second, the molecular quantum dynamics is evaluated using mean field theory, specifically it is assumed that the molecular wavefunction can be approximately written as a product of wavefunctions of the individual molecules. While the validity of these approximations is not obvious, they were found to compare favorably, in the weak excitation limit, with a fully quantum calculation of a similar model in the one-exciton approximation. While the strong coupling character of our simple model is clearly established and can be clearly identified in the numerical results for transmission and reflection, a significant observation is the early time slowdown of the dissociation dynamics for when the system is driven at the upper or lower polariton frequencies.

For a qualitative understanding of this observation consider the ideal case where the Condon approximation is valid so that the electronic excitation does not change the vibrational state. Let an N-molecules system start in its ground electronic/vibrational state, $\left|\Psi_g\right\rangle = \prod_{j=1}^{N}\left|g_j\right\rangle\left|\chi_{gj}\right\rangle$ where $\left|g_j\right\rangle$ and $\left|\chi_{gj}\right\rangle$ denote the ground electronic state and the corresponding vibrational state of molecule $j$. A sudden single photon absorption creates an excitation (an exciton outside and a polariton inside the cavity) that involves a molecular bright vibronic state of the form

$$\left|\Psi_x\left(t=0\right)\right\rangle = \frac{1}{\sqrt{N}}\left(\sum_j\left|X_j\right\rangle\right)\prod_j\left|\chi_{gj}\right\rangle, j = 1,\dots,N, \qquad (17)$$

where, using $\left|x_j\right\rangle$ for the electronically excited molecular state,

$$\left|X_j\right\rangle = \left|x_j\right\rangle\prod_{k\neq j}\left|g_k\right\rangle. \qquad (18)$$

The ensuing nuclear dynamics can be described by a wavefunction of the form

$$\left|\Psi_x\left(t\right)\right\rangle = \frac{1}{\sqrt{N}}\sum_j\left|X_j\right\rangle\left|\chi_j\left(t\right)\right\rangle\prod_{k\neq j}\left|\chi_{gk}\right\rangle, \qquad (19)$$

where $\chi_j\left(t\right)$ describes the time evolution of a nuclear state that moves on the excited potential surface of molecule $j$ starting from $\chi_j\left(t=0\right) = \chi_{gj}$. Outside the cavity this leads an evolution similar to the separation of nuclear trajectories in a post-conical-intersection evolution as nuclear states $\left|\chi_j\left(t\right)\right\rangle\prod_{k\neq j}\left|\chi_{gk}\right\rangle$ moving on different potential surfaces fall out of overlap with each other. In the presence of coupling to a cavity mode the situation is more complicated. Denoting the ground and singly excited cavity states by $\left|0\right\rangle$ and $\left|1\right\rangle$ respectively, the zero-order (in the molecule-cavity interaction) states $\left|\Psi_g\right\rangle\left|1\right\rangle$ and $\left|\Psi_x\left(t=0\right)\right\rangle\left|0\right\rangle$ couple to form a collective polariton whose vibrational component is $\prod_j\left|\chi_{gj}\right\rangle$ with a polaritonic splitting of order $\sqrt{N}$. If the nuclear motion proceeds according to Eq. (19) this collective nature of the coupling is lost because each component of (19) when combined with the cavity state, namely the state $\left|0\right\rangle\left|X_j\right\rangle\left|\chi_j\left(t\right)\right\rangle\prod_{k\neq j}\left|\chi_{gk}\right\rangle$ is now coupled to an excited cavity state dressed by a different vibrational state $\left|1\right\rangle\prod_{j=1}^{N}\left|g_j\right\rangle\left|\chi_j\left(t\right)\right\rangle\prod_{k\neq j}\left|\chi_{gk}\right\rangle$. Once the overlap between different vibrational states $\left|\chi_j\left(t\right)\right\rangle\prod_{k\neq j}\left|\chi_{gk}\right\rangle$ decays, the Rabi splitting loses its collective $\sqrt{N}$ order and becomes of order 1, namely of the magnitude of a single molecule coupling to the radiation field.

This state of affairs implies that the separation of nuclear trajectories implied by Eq. (19) now has an additional energetic cost associated with the loss of the collective Rabi splitting. The detailed implications of this observation for our present dissociation dynamics model will be considered elsewhere, and here we provide a qualitative argument based on a recent analysis by Cui and Nitzan[30] who have used a simplified truncated basis model to investigate the dynamic consequences of this situation. In that paper it was argued that in contrast to previously discussed polariton decoupling models, the loss of collectivity due to nuclear motion following electronic polariton excitation implies a *local* distortion of the nuclear potential surface: a local maximum in the nuclear surface associated with excitation of the lower polariton (where energy increases as the collective Rabi splitting is lost) and a local minimum the surface associated with excitation to the upper polariton. In both cases (see a sketch in Fig. 6) this implies the emergence of a local barrier that causes a transient slowing down of the nuclear propagation, although for the upper polariton this slowing-down appears to be preceded by a transient acceleration, which may explain the difference between the dynamics of the upper and lower polariton seen in Fig. 5. Of course, this handwaving argument should not be taken as a basis for quantitative understanding but suggests that transition from collective to individual molecular behavior induced by the nuclear motion following molecular polariton excitation in an optical cavity can have interesting dynamical consequences and may be the reason for the transient slowdown observed in our simulations.

## 5. Conclusion

In this work we considered photodissociation dynamics of an ensemble of diatomic molecules under strong coupling conditions, using a semiclassical model where the electromagnetic field is described by the Maxwell equations and the molecular system is treated using a self-consistent mean field approximation. This yields a set of coupled Maxwell-Schrödinger equations with the molecular ro-vibrational degrees of freedom explicitly included. The combined effect of collective molecular excitation (that is manifested in a considerable polariton energy shift) and the individual nature of molecular nuclear dynamics (namely, nuclear motion responds to the excitation state of individual molecules) is shown to have significant consequences for the dissociation dynamics following polariton excitation. Specifically, the rate of nuclear separation following polariton formation is initially slower than the corresponding process following excitations into spectral regimes where a polariton is not formed. A rationalization of the observed effect proposed was based on the recognition that nuclear separation in individual (excited) molecules is essentially a dephasing process that tends to destroy the polariton and therefore has an energy cost. The resulting local activation barrier leads to a local (therefore transient slow-down of the nuclear separation dynamics. Further consequences of the conflicting effects of collective excitation and individual molecular motions should be expressed in the presence of thermal interactions and will be explored in the future.

## 6. Acknowledgements

This work is supported by the Air Force Office of Scientific Research under grant No. FA9550-22-1-0175 (MS), by the U.S. National Science Foundation under Grant No. CHE1953701 (AN), and by The U.S. Department of Energy, Office of Science, Basic Energy Sciences, Chemical Sciences, Geosciences, and Biosciences Division, under Award No. DE-SC0019397 (JS).

**FIGURES**

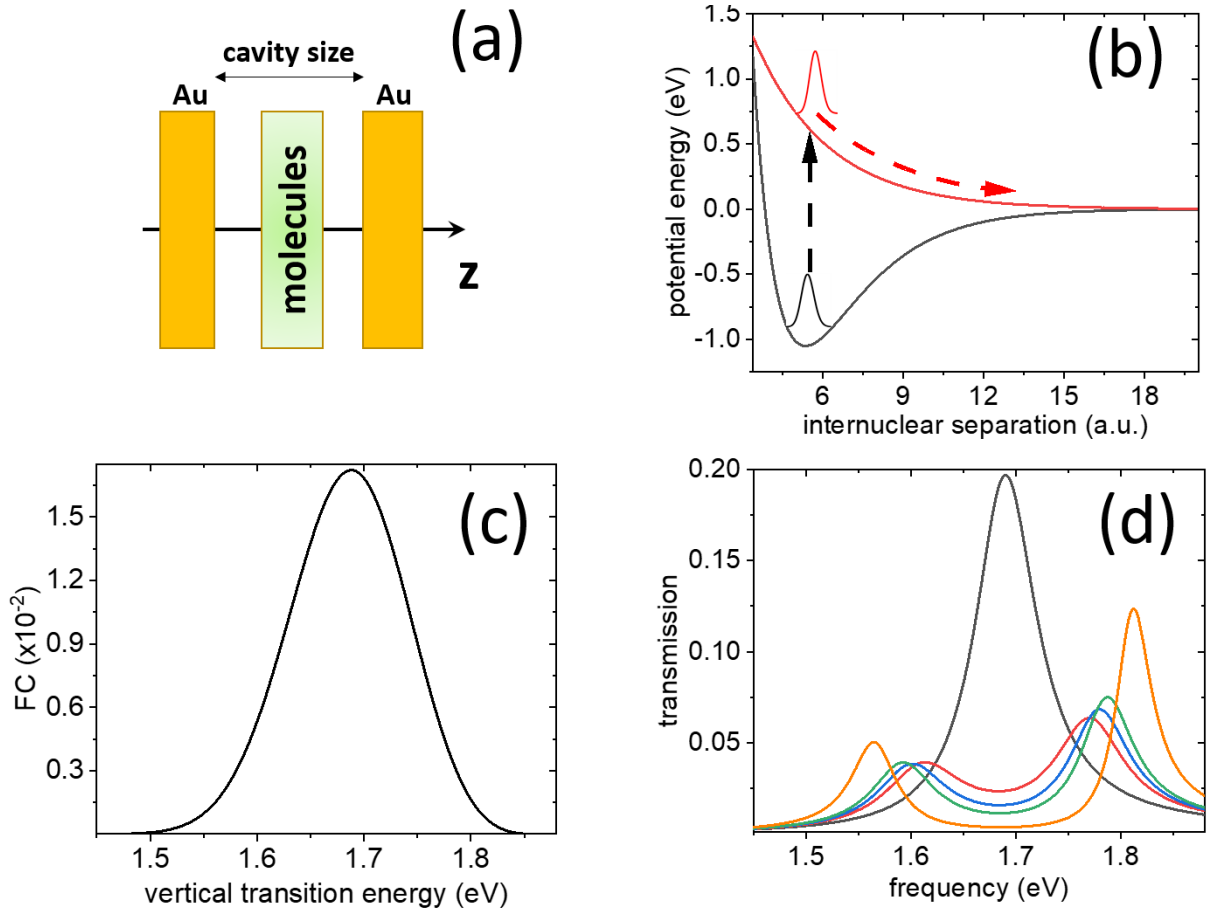

**Figure 1**. Schematic setup of a 1D optical cavity with a molecular slab inside is shown in panel (a). Panel (b) shows electronic potential energy surfaces (see Eqs. (11) and (12)). Franck-Condon factors, $FC$, as defined in Eq. (14) are plotted in panel (c) as functions of the transition energy. Panel (d) shows linear transmission as a function of the incident frequency: black line shows data for an empty cavity, red line is with a molecular slab with the number density of $6 \times 10^{25}$ m⁻³, blue line is for $8 \times 10^{25}$ m⁻³, green line shows data for $10^{26}$ m⁻³, and dark orange line shows results for $2 \times 10^{26}$ m⁻³. The cavity size is 305 nm. The thickness of the mirrors is 40 nm.

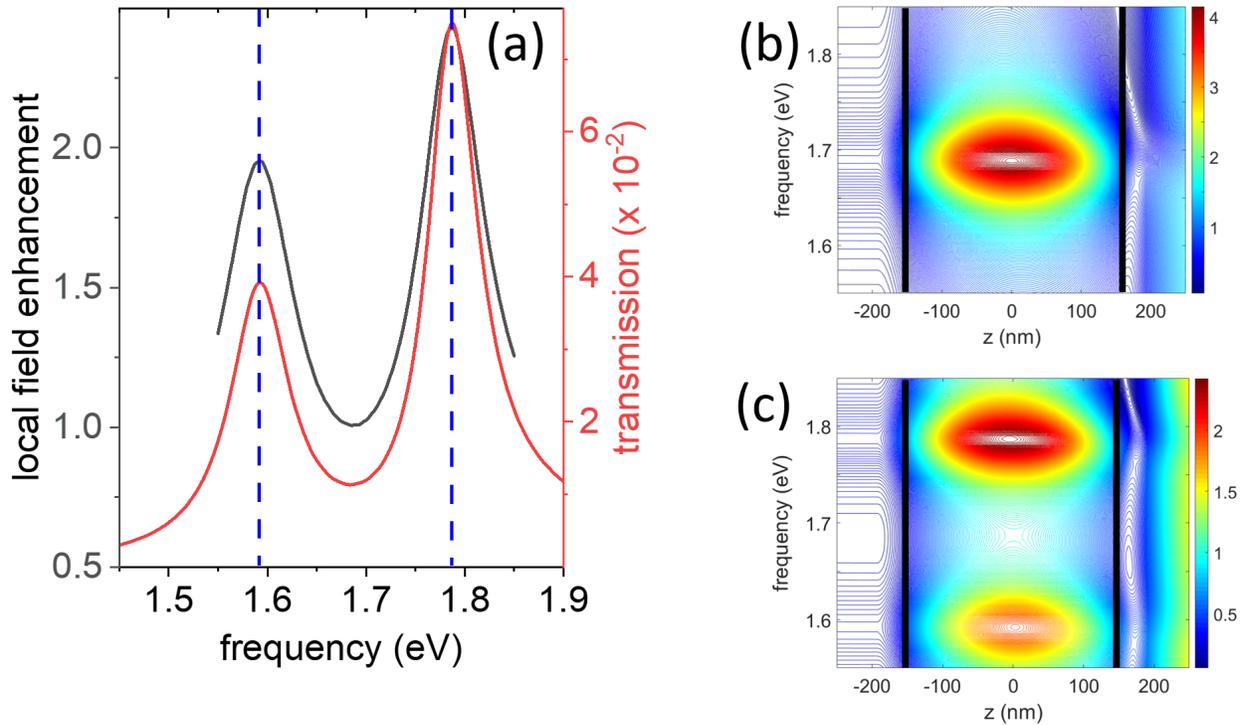

**Figure 2**. Panel (a) shows the local field enhancement (black line, left vertical scale) and the linear transmission (red line, right vertical scale) as functions of the incident frequency. Two vertical dashed lines indicate polaritonic frequencies. The cavity parameters are the same as in Fig. 1d. The molecular number density is $10^{26}$ m$^{-3}$. Panels (b) and (c) show spatial (horizontal axis) and frequency (vertical axis) distributions of the local field enhancement in the empty cavity (b) and in the cavity with the molecular slab (c). Parameters of the cavity and molecules are the same as in panel (a). Vertical black lines in panels (b) and (c) show positions of the mirrors.

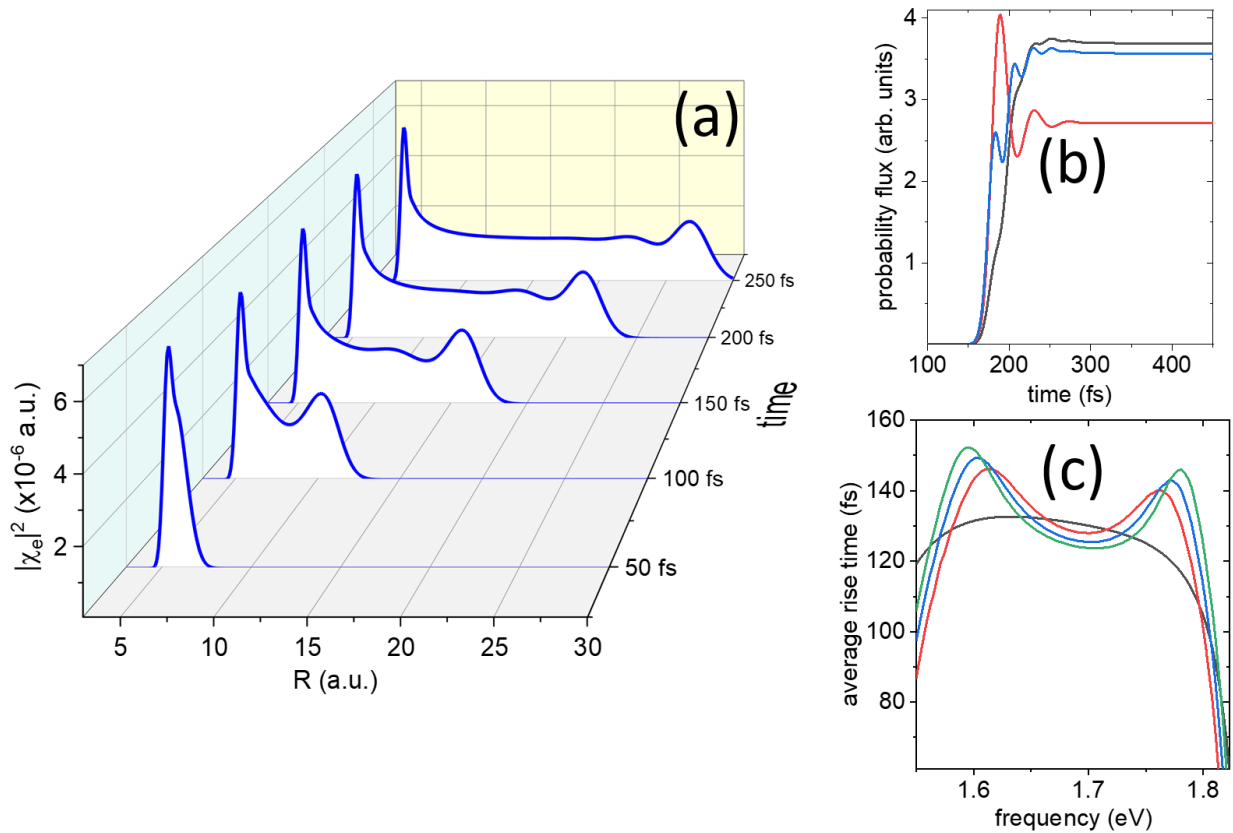

**Figure 3**. Time-dynamics of the excited ro-vibrational wavefunction (panel (a)) as a function of $R$ and time. Calculations are performed for the driving field at the molecular transition frequency (1.69 eV) for the number density of $10^{26}$ m$^{-3}$. Panel (b) shows the probability flux in the excited electronic state as a function of time calculated at $R = 15$ a.u.: black line is for the driving frequency corresponding to the lower polariton, red line is for the molecular transition frequency, and blue is the for upper polariton. Calculations are performed for the number density of $10^{26}$ m$^{-3}$. Panel (v) shows the average rise time (see text) of the probability flux as a function of the driving frequency: black line shows results for the molecular slab in vacuum (these do not depend on the molecular density), red line is for the number density of $6 \times 10^{25}$ m$^{-3}$, blue line is for $8 \times 10^{25}$ m$^{-3}$, green line shows data for $10^{26}$ m$^{-3}$. The other parameters are the same as in Fig. 2a.

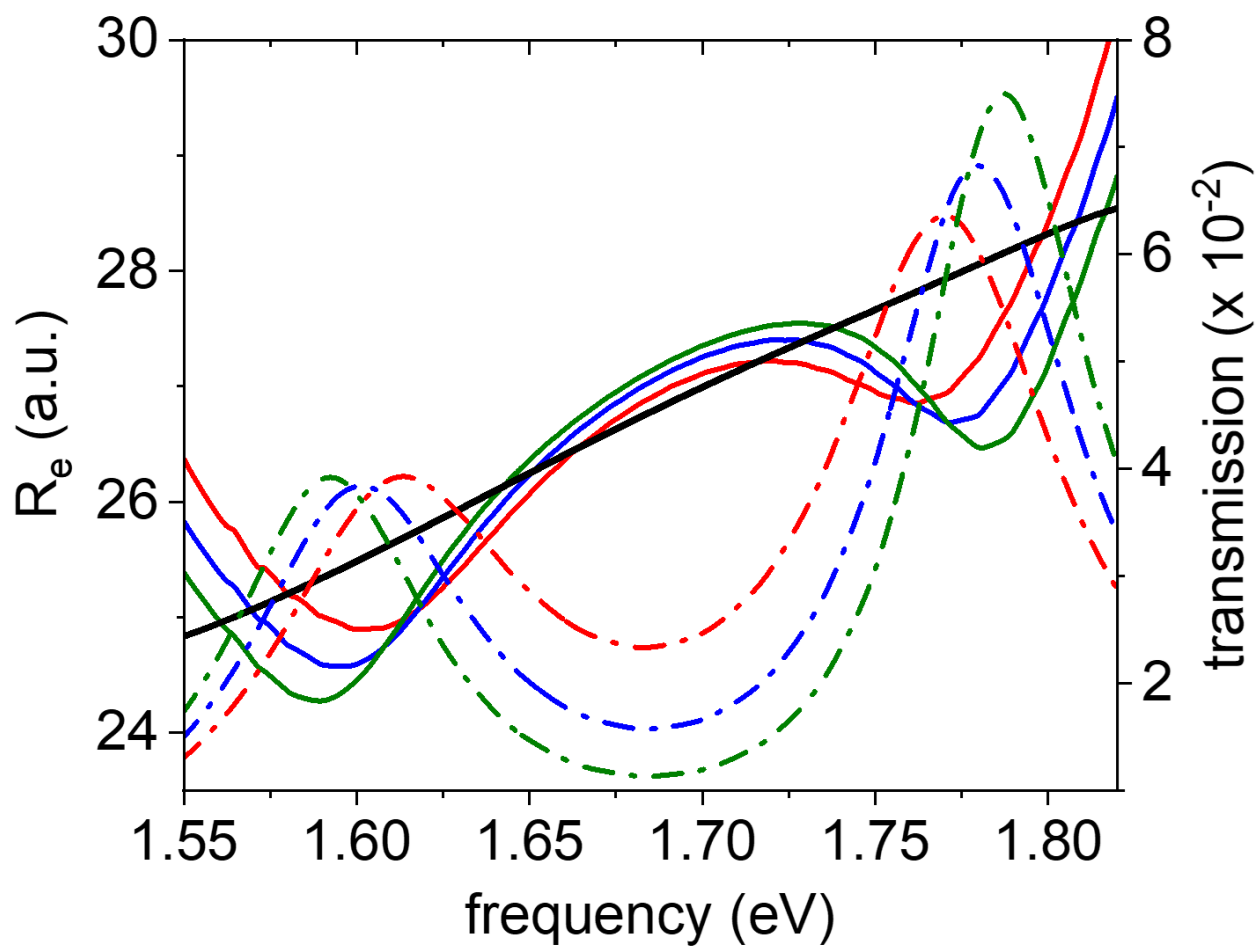

**Figure 4**. Ensemble averaged internuclear distance in the excited electronic state (left vertical axis) calculated at 450 fs after the CW field is turned on. Black line shows results for the molecular slab in vacuum (no cavity, results do not depend on the molecular concentration). Red line shows data for the number density of $6 \times 10^{25}$ m$^{-3}$, blue line is for $8 \times 10^{25}$ m$^{-3}$, and green line shows data for $10^{26}$ m$^{-3}$. Dash-dotted lines show linear transmission (right vertical scale) for the same values of the number density using the same color scheme. Other parameters are the same as in Fig. 3.

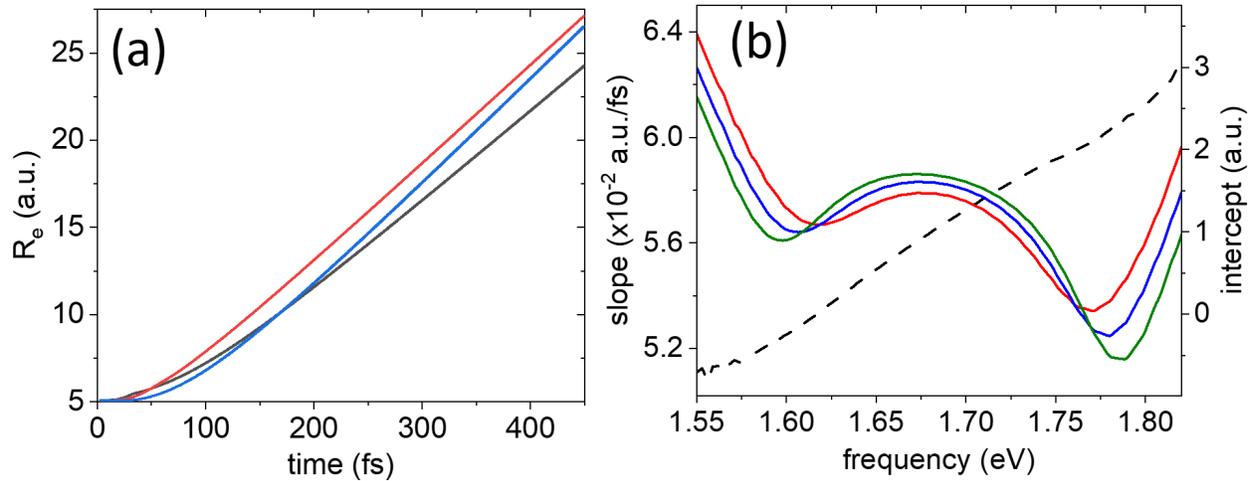

**Figure 5**. Dissociation dynamics under resonant conditions. The weighted ensemble averaged internuclear distance is shown in panel (a) as a function of time driven at the lower polariton frequency (black line), molecular transition frequency (red line), and the upper polariton frequency (blue line). Simulations are performed inside the resonant cavity at the molecular concentration of $10^{26}$ m$^{-3}$. Time 0 corresponds to the beginning of the time propagation when CW field is gradually turned on for 5 fs. Panel (b) shows results of linear regression calculations plotting the slope (left vertical axis) and the intercept (right vertical axis) as functions of the driving frequency. Calculations are performed for molecules in the resonant cavity at the number density of $6\times10^{25}$ m$^{-3}$ (red), $8\times10^{25}$ m$^{-3}$ (blue), and $10^{26}$ m$^{-3}$ (green). The dashed black line shows the intercept (it does not depend on the number density of molecules). Other parameters are the same as in Fig. 3.

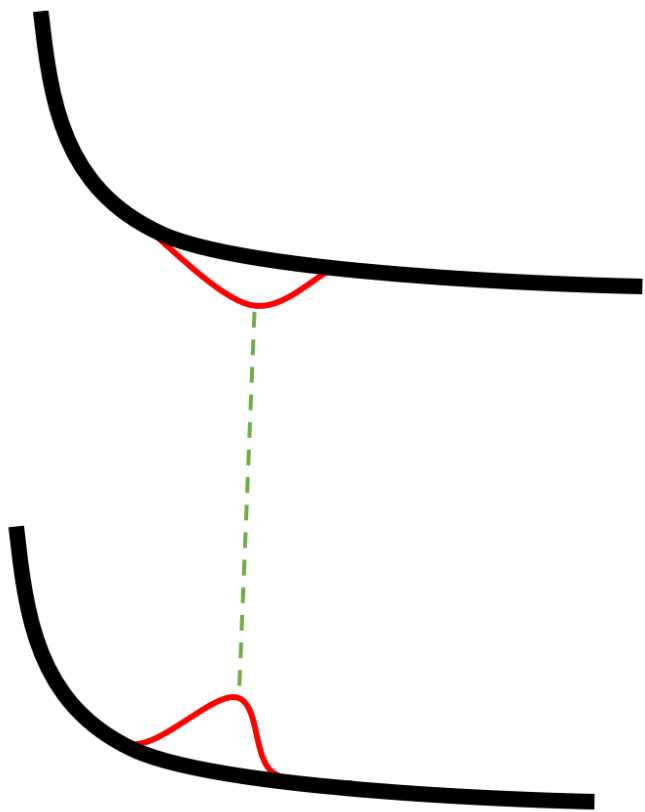

**Figure 6**. Schematic energy diagram that illustrates the Rabi splitting vs. internuclear separation during dissociation process.